# Rapid yet accurate first principle based predictions of alkali halide crystal phases using alchemical perturbation

Alisa Solovyeva[1] and O. Anatole von Lilienfeld[1, 2, ∗]

[1]*Institute of Physical Chemistry and National Center for Computational Design and Discovery of Novel Materials, Department of Chemistry, University of Basel, Klingelbergstrasse 80, 4056 Basel, Switzerland*
[2]*General Chemistry, Free University of Brussels, Pleinlaan 2, 1050 Brussel, Belgium*
(Dated: May 25, 2016)

We assess the predictive power of alchemical perturbations for estimating fundamental properties in ionic crystals. Using density functional theory we have calculated formation energies, lattice constants, and bulk moduli for all sixteen iso-valence-electronic combinations of pure pristine alkali halides involving elements $A \in$ {Na, K, Rb, Cs} and $X \in$ {F, Cl, Br, I}. For rock salt, zincblende and cesium chloride symmetry, alchemical Hellmann-Feynman derivatives, evaluated along lattice scans of sixteen reference crystals, have been obtained for all respective 16×15 combinations of reference and predicted target crystals. Mean absolute errors (MAE) are on par with density functional theory level of accuracy for energies and bulk modulus. Predicted lattice constants are less accurate. NaCl is the best reference salt for alchemical estimates of relative energies (MAE < 40 meV/atom) while alkali fluorides are the worst. By contrast, lattice constants are predicted best using NaF as a reference salt (MAE < 0.5Å), yielding only semi-quantitative accuracy. The best reference salt for the prediction of bulk moduli is CsCl (MAE < 0.4×$10^{11}$ dynes/cm$^2$). Alchemical derivatives can also be used to predict competing rock salt and cesium chloride phases in binary and ternary solid mixtures with CsCl. Alchemical predictions based on dispersion corrected density functional theory with pure RbI as a reference salt reproduce reasonably well the reversal of the rock salt/cesium chloride stability trend for binary $(AX)_{1-x}CsCl_x$ as well as for ternary $(AX)_{0.5-0.5x}(BY)_{0.5-0.5x}CsCl_x$ mixtures.

## I. INTRODUCTION

The accurate prediction of crystal structures represents a crucial aspect in our understanding of phase diagrams. The crystal structure prediction blind test gauges the performance of the state of the art in the field on a regular basis [1, 2]. Organic crystals are particularly challenging due to the necessity of accurate inter and intra-molecular potentials including many-body van der Waals contributions [3, 4]. In addition to a sufficiently accurate potential energy model, spatial degrees of freedom need to be sampled in an efficient manner to locate competing polymorphs. Various methods have been introduced to accomplish the latter [5–8] All these methods succeed in finding local, and global, potential or free energy minima of competing phases for any given material. While great progress has been made in the context of predicting pure and pristine phases, predicting energies and structures of doped materials, solid mixtures, and co-crystals represents an even more complex challenge. Furthermore, when it comes to virtual materials design, not only configurational but also compositional degrees of freedom have to be taken into account, as recently exemplified by Marques, Botti and co-workers [9]. In this study, we have investigated the applicability of "alchemical" coupling in order to rapidly estimate stability, structure, and properties of competing crystal phases of varying composition *without* having to perform brute-force screening.

"Alchemical coupling" refers to adiabatically connecting external potentials of two materials in a way that typically includes a continuous variation in nuclear charges. As such, the coupling paths have no correspondence in reality, and hence we refer to them as "alchemical" [10]. All properties which are thermodynamic state functions can be coupled using arbitrary interpolation functions between the two end points. Alchemical paths are common in force-field based free energy calculations [11, 12], and have found various applications such as virtual drug screening [13, 14], or determination of eutectic mixtures of heat transfer fluid candidates [15]. They have been less common in quantum mechanics, despite their early proposition in 1962 [16]. An early effort is a 1975 study on continuous changes of electronic valence into Rydberg states [17]. By now they are no longer unusual and have become widely spread for predicting the effect of compositional changes on a broad variety of properties, including energies, free energies, nuclear quantum effects, and electronic properties of systems in gas, liquid and solid phase [10, 18–39]. For more details and references, we refer to two recent reviews [40, 41]. As long as it is sufficiently accurate, any gradient based exploration campaign is dramatically more effective than brute force screening or discrete alternatives, be it using self-consistent field procedures or extended molecular dynamics trajectories. In this study, we have systematically assessed the performance of alchemical coupling for the prediction of properties in a well defined class of materials: We studied alchemical coupling of alkali halide (AX) crystals, often used to benchmark novel crystal structure approaches [42]. We chose this class of com-pounds because they represent an appealing compromise: They have a non-trivial degree of chemical diversity, yet their dominant nature of cohesion is simple, solely due to ionic bonding. As such, we consider them to represent an important benchmark: If alchemical derivatives (or any other approach for that matter) already failed to describe alkali halides one would hesitate to proceed to more challenging crystals which entail, for example, also covalent or intermolecular binding.

---

∗ anatole.vonlilienfeld@unibas.ch

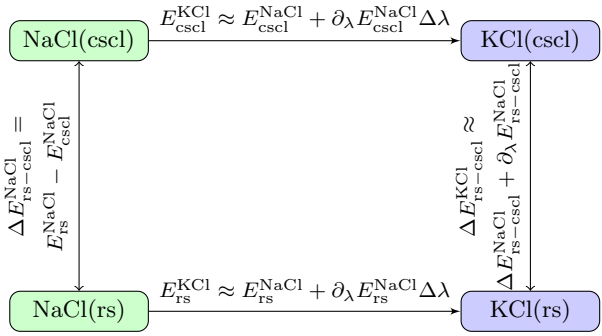

FIG. 1. Thermodynamic cycle used to alchemically (horizontal arrows) predict the rocksalt (bottom)/cscl (top) energy difference of KCl (right) using NaCl (left) as a reference.

This work is organized as follows: In Sec. II, we briefly summarize the general theoretical basis for first order alchemical derivatives within density functional theory (DFT). Computational details are discussed in Sec. III, followed by results for pure alkali halides in Sec. IV C. In Sec. IV D and Sec. IV E we analyze the performance of the first order alchemical derivatives for binary and ternary alkali halide mixtures. In Sec. V we summarize this study and provide concluding remarks.

## II. THEORY

We couple any two iso-electronic crystals, consisting of initial reference system $r$ and target system $t$ with a global Hamiltonian, linear in coupling parameter $\lambda$,

$$\hat{H}(\lambda) = \hat{H}^r + \lambda(\hat{H}^t - \hat{H}^r). \tag{1}$$

Here $0 \leq \lambda \leq 1$, and $\hat{H}$ refers to the total electronic Hamiltonian, i.e. including nuclear-nuclear repulsion. Also, we consider only "vertical" alchemical changes, i.e. initial and final crystal structures always have the same number of atoms located at the exact same points in space in the same crystal structure. Note, that non-linear interpolations are also possible for $\hat{H}(\lambda)$ [32], but have not been explored in this study. For alchemical changes involving elements from different rows in the periodic table, we can easily restrict ourselves to changes which are iso-electronic in valence electrons only, and account for the core electron's changes through interpolation of their effective core (or pseudo-)potential. All alchemical changes investigated in this paper only include moves going up or down the same column in the periodic table (e.g. Na → Cs, or Br → F). First order derivatives with respect to such alchemical changes have just recently been shown to have superior predictive power in the case of covalent bonding in small molecules [38]. The first order derivative according to Hellmann–Feynman [43, 44] corresponds for alchemical coupling within DFT to [32],

$$\left.\frac{\partial E}{\partial \lambda}\right|_{\lambda=0} = \langle \Psi^r | \hat{H}^t - \hat{H}^r | \Psi^r \rangle = E^t[n^r] - E^r[n^r]$$

$$= \int d\mathbf{r}\, n^r(\mathbf{r})(v^t(\mathbf{r}) - v^r(\mathbf{r})) \tag{2}$$

where, $\Psi$, $n$ and $E[n]$ are the corresponding unknown electronic wavefunction, density, and energy functional, respectively. And $v$ is the known external potential. Note that this is identical with the energy expression in first order perturbation theory, when using $\partial_\lambda H$ as the perturbing Hamiltonian. We use Eq. 2 to estimate the total energy of the target in a first order Taylor expansion using only the electron density of the reference system. The predicted energy, $E^p$, thus has the form,

$$E^t \approx E^p = E^r[n^r] + \Delta\lambda \partial_\lambda E_\lambda \bigg|_{\lambda=0}, \tag{3}$$

and simplifies for $\Delta\lambda = 1$ to $E^p = E^t[n^r]$. Note that in general $E^t[n^r]$ is a very poor model of $E^t[n^t]$, espe-cially when it comes to the prediction of absolute energies. However, it turns out that usually this is mostly due to a constant shift in the off-set, shape and location of $E^t[n^r]$ as a function of lattice constant agree very well with $E^t[n^t]$. We can exploit this finding when we restrict ourselves to predicting alchemical changes in *relative* energies, rather than absolute energies. This is not a severe restriction since the latter are arbitrary within pseudopotential based calculations anyhow, and since they, maybe more importantly, hardly matter for most of the common physical and chemical processes.

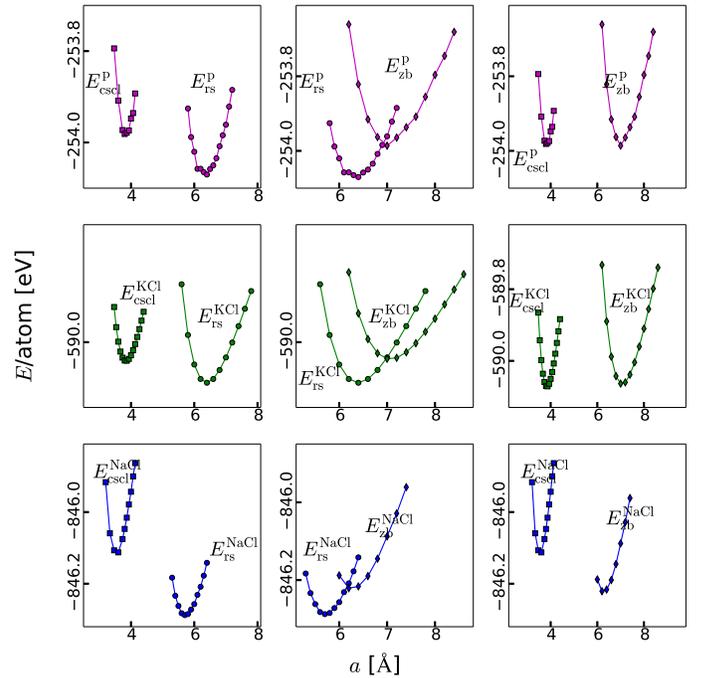

FIG. 2. Calculated absolute total potential energies as a function of lattice parameter for rocksalt (rs), cesium chloride (cscl), and zincblende (zb) phases of NaCl and KCl. Top panels correspond to alchemical predictions of KCl using NaCl as a reference, and according to Eq. (3). Mid and Bottom panels correspond to DFT/PBE calculations of KCl and NaCl, respectively.

| $a$ [Å] | F | Cl | Br | I |
|---|---|---|---|---|
| | | rs | | |
| Na Exp., $T_a$ | 4.62 | 5.64 | 5.97 | 6.47 |
| Exp., $T_0$ | 4.61 | 5.60 | 5.93 | 6.41 |
| DFT | 4.80 | 5.70 | 6.00 | 6.60 |
| K Exp., $T_a$ | 5.34 | 6.29 | 6.60 | 7.07 |
| Exp., $T_0$ | 5.31 | 6.25 | 6.54 | 6.99 |
| DFT | 5.40 | 6.40 | 6.80 | 7.20 |
| Rb Exp., $T_a$ | 5.65 | $6.59^a$ | 6.89 | 7.34 |
| Exp., $T_0$ | 5.59 | 6.53 | 6.82 | 7.26 |
| DFT | 5.80 | 6.80 | 7.00 | 7.60 |
| | rs | | cscl | |
| Cs Exp., $T_a$ | $6.02^a$ | 4.12 | 4.30 | 4.57 |
| Exp., $T_0$ | | | $4.23^b$ | $4.51^b$ |
| DFT | 6.20 | 4.20 | 4.40 | 4.67 |

TABLE I. Experimental [45] and calculated (DFT/PBE) lattice constants for all pure alkali halides considered in this study. Experimental values correspond to $T_a$=298 K and $T_0$=0 K. Values marked with $^a$ were obtained at T=293 K, and $^b$ denotes values from Ref. [46].

| $\Delta E_{\text{rs-cscl}}$/atom [meV] | F | Cl | Br | I |
|---|---|---|---|---|
| Na Ref. [47] | −142 | −121 | −117 | −91 |
| DFT | −163 | −175 | −181 | −194 |
| K Ref. [47] | −108 | −90 | −90 | −75 |
| DFT | −100 | −84 | −82 | −89 |
| Rb Ref. [47] | −94 | −68 | −61 | −49 |
| DFT | −94 | −60 | −58 | −57 |
| Cs Ref. [47] | −105 | −65 | −56 | −45 |
| Ref. [48] | −110 | −50 | −40 | −30 |
| DFT | −106 | −40 | −36 | −30 |

TABLE II. Calculated differences between equilibrium total energies, $\Delta E_{\text{rs-cscl}}$, obtained in this work (DFT/PBE) or from the literature (Ref).

| $B$ [$10^{11} \frac{\text{dynes}}{\text{cm}^2}$] | F | Cl | Br | I |
|---|---|---|---|---|
| | | rs | | |
| Na Exp. 1 | $5.14^d$ | $2.66^d$ | $2.26^d$ | $1.79^e$ |
| Exp. 2 | 4.85 | 2.49 | 2.04 | 1.61 |
| Exp. 3 | 4.60 | 2.10 | | |
| DFT | 4.08 | 2.29 | 2.07 | 1.43 |
| K Exp. 1 | $3.42^d$ | $1.97^f$ | | $1.27^f$ |
| Exp. 2 | 3.17 | 1.81 | 1.52 | 1.20 |
| Exp. 3 | | 1.76 | 1.46 | 1.13 |
| DFT | 2.98 | 1.61 | 1.21 | 1.10 |
| Rb Exp. 1 | $3.01^g$ | $1.87^d$ | $1.60^d$ | $1.31^d$ |
| Exp. 2 | 2.77 | 1.63 | 1.37 | 1.10 |
| Exp. 3 | | 1.48 | 1.28 | 1.03 |
| DFT | 2.27 | 1.17 | 1.23 | 0.85 |
| | rs | | cscl | |
| Cs Exp. 1 | | | $1.84^h$ | $1.44^h$ |
| Exp. 2 | 2.50 | 1.82 | 1.58 | 1.26 |
| DFT | 1.87 | 1.44 | 1.19 | 0.99 |
| | | cscl | | |
| K Exp. 3 | | 1.728 | 1.117 | 0.987 |
| DFT | 3.37 | 1.71 | 1.50 | 1.18 |
| Rb Exp. 3 | | | 1.148 | 1.105 |
| DFT | 2.77 | 1.65 | 1.39 | 1.02 |

TABLE III. Experimental and calculated (DFT/PBE) bulk moduli. Experimental values are obtained at T=4.2 K ($^d$[49], $^e$[50], $^f$[51], $^g$[52], $^h$[46]) (Exp. 1) and as an average of room temperature values from the Landolt–Börnstein tables (Exp. 2)[53]. In Ref. [54] bulk moduli are determined spectroscopically (Exp. 3).

Fig. 1 illustrates the thermodynamic cycles one can use to make alchemical predictions of changes in relative energies. Green boxes correspond to reference salts (example NaCl) and blue to target structures (example KCl). Note that any other iso-valence-electronic combination of reference and target crystal could have been used, even including binary, ternary, quaternary, etc. alkali halide mixtures. Thus, knowing the electron density of a single reference compound holds promise to access a vast range of iso-valence-electronic compounds via alchemical Hellmann–Feynman derivatives.

For predictions of relative energies between phases, such as cesium chloride (cscl) versus rock salt structure (rs), the first order derivative becomes,

$$\begin{aligned} E^t_{\text{cscl-rs}} &\approx E^r_{\text{cscl-rs}} + \Delta\lambda\partial_\lambda E^r_{\text{cscl-rs}} \\ &= E^r_{\text{cscl-rs}} + E^t_{\text{cscl}}[n^r_{\text{cscl}}] - E^t_{\text{rs}}[n^r_{\text{rs}}] + E^r_{\text{rs}} - E^r_{\text{cscl}} \\ &= E^t_{\text{cscl}}[n^r_{\text{cscl}}] - E^t_{\text{rs}}[n^r_{\text{rs}}] \quad = \quad E^p_{\text{cscl}} - E^p_{\text{rs}} \end{aligned} \quad (4)$$

In order to obtain estimates of meaningful relative energies, we report predicted relative energies evaluated at those lattice constant values of each reference crystal which correspond to the minima of the respective predicted energy curves. For example, $E^t_{\text{cscl}}[n^r_{\text{cscl}}]$ is evaluated using the reference electron density in cscl structure obtained at that lattice constant value which minimizes $E^p$ in cscl structure. Conversely, $E^t_{\text{rs}}[n^r_{\text{rs}}]$ is evaluated using the reference electron density in rs structure obtained at that lattice constant value which minimizes $E^p$ in rs structure.

Higher order derivatives in the energy Taylor expansion could possibly increase the accuracy of alchemical predictions [57]. Their convergence, however, should not be taken for granted [38]. The most straightforward way to include them is by finite difference. In practice, however, it is difficult to go beyond 2nd order due to numerical noise. It would go beyond the scope of this study to also include higher order effects. Furthermore, calculation of higher order derivatives lead to an increase in computational cost, which is why first order derivatives should be fully explored first.

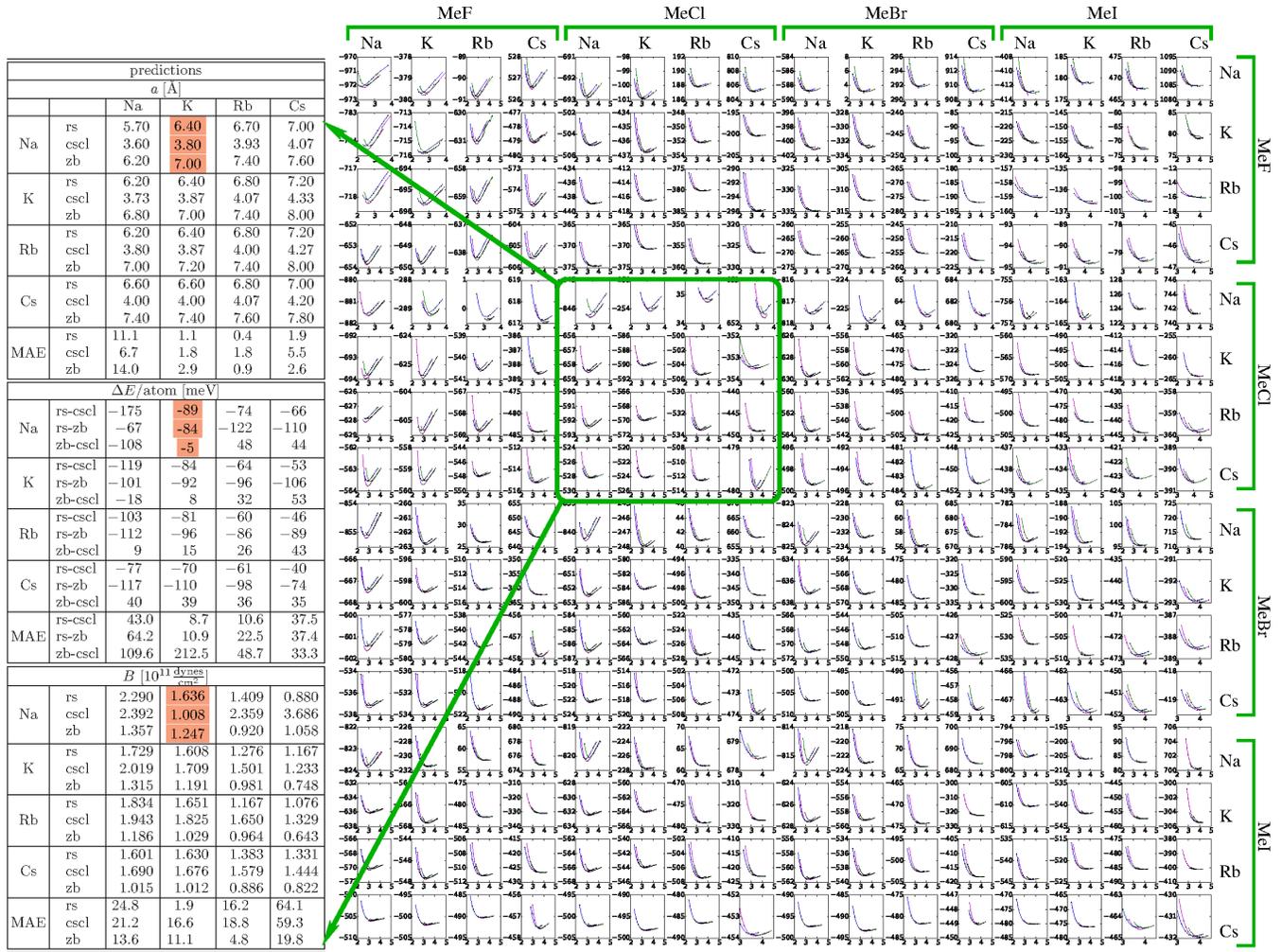

FIG. 3. RIGHT: Table containing alchemically predicted scans of energies [eV] as a function of interatomic distance [Å] for all possible 16×15 pure alkali halide couplings considered here within. Diagonal plots correspond to true DFT/PBE results used as reference to alchemically predict all off-diagonal plots in the same column. Thus, rows and columns indicate reference and target salts, respectively. In each plot there is a pink, green, and blue line corresponding to rs, cscl, and zb phase, respectively. LEFT: Zoom-in for all chlorides. The table lists relative energies, equilibrium lattice constants, and bulk moduli extracted from corresponding alchemical prediction scans. Again, the values in the diagonal elements correspond to DFT reference numbers, and columns indicate target chlorides (for convenience Cl symbols have been dropped), while rows correspond to ACl reference salts. Lattice constants and bulk moduli have been obtained by fitting calculated data to the BM equation of state[55, 56]. Energy difference between two minima corresponding to two different symmetries. Numbers highlighted in orange have been extracted from energy curves already presented in Fig. 2.

## III. COMPUTATIONAL DETAILS

All calculations have been performed with plane wave/pseudopotential based DFT [58, 59], as implemented in the CPMD code [60]. First order Hellmann–Feynman derivative based estimates for target salts have been evaluated using CPMD's RESTART files containing the electron density of the reference salt. After one iteration the self-consistent field cycle is aborted, and $E^t[n^r]$ is calculated.

Regarding the exchange–correlation potential, PBE[61] and LDA[62, 63] functionals have been used. Goedecker–Teter–Hutter (GTH) pseudopotentials[64, 65] and a plane-wave cutoff of 250 Ry have been used throughout. For halogens ($X$) and alkali atoms ($A$), we employed pseudopotentials with effective nuclear charges of seven, and nine, respectively. Examination of alkali metals with nuclear charge equals one indicated poor performance for most of the salts with slow convergence and strong oscillations of the total energy as a function of cell size. For this reason, we have excluded lithium from this study. The wavefunction convergence criterion has been set to $10^{-7}$ Ha. Γ-point only (no $k$-point sampling) has been used. The rs and Zincblende (zb) crystal structures were modeled by 64 atoms, for cscl we used a unit cell containing 54 atoms. The Birch–Murnaghan (BM) isothermal equation of state[55, 56] has been employed to fit data points and estimate bulk moduli of predicted as well as reference curves. We also added the widely spread force-field like two-body dispersion energy contributions [3, 66–68] to the DFT/PBE reference energies, as well as to the $E^p$.

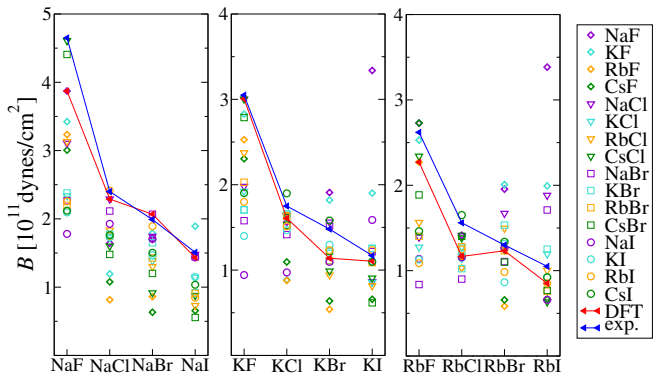

FIG. 4. Alchemical predictions of bulk moduli of alkali halides in rs structure. Abscissa indicates target crystal, and legend indicates reference crystal. All the data has been extracted from the predictions shown in Fig. 3.

$$E^{\rm DFT+D2} = E^{\rm DFT} + E^{\rm D2} \quad (5)$$

$$= E^{\rm DFT} - s_6 \sum_{i=1}^{N-1} \sum_{j=i+1}^{N} f(r_{ij}) \frac{C_6^{ij}}{r_{ij}^6}, \quad (6)$$

where $s_6$ is the scaling factor, $N$ - the number of atoms, $C_6^{ij}$ - dispersion coefficient for an atom pair, $r_{ij}$ - interatomic distance, and $f(r_{ij})$ is a damping function. The need to calculate $E^{\rm D2}$ arises, since pure DFT/PBE incorrectly predicts the rs to be more stable than cscl for CsCl, CsBr, and CsI [48]. The dispersion correction has been obtained following Ref. [48], where $C_6$ and $R_0$ parameters were taken from Ref. [68] for all elements but for Cs which was taken from Ref. [48]. Note that this dispersion correction is added to the alchemical prediction *a posteriori*. This is more straightforward than including it through alchemy since there is no explicit $\lambda$ dependence in this dispersion correction. At this point we note that dispersion coefficient based corrections could have also been obtained "on-the-fly" with minimal empirical effort following the procedure proposed in Ref. [69].

## IV. RESULTS AND DISCUSSION

### A. Performance of DFT for describing alkali halides

We have calculated DFT/PBE lattice constants (Tab. I), energy differences (Tab. II), and bulk moduli (Tab. III) for all alkali halides in their lowest energy phase. Available experimental data and previously performed theoretical results by others are also listed for comparison.

Theoretical values of lattice parameters in Tab. I differ from experiment by 1.5-2% (0.1Å or more) for the majority of $AX$. The highest deviation occurs for RbI: 4.5%. DFT/PBE systematically overestimates the experimental value at T=0 K, even though being calculated for static structures. This behavior of the PBE functional has also been observed in previous theoretical studies, e.g. see Ref. [70]. The overall performance of DFT lattice parameters, however, is satisfying. In particular, all trends are in perfect agreement with experiment.

In Tab. II we report DFT/PBE energy differences between rs and cscl calculated in the present work for all pure $AX$ crystals, along with previous theoretical results by others [47, 48]. Note that in Ref. [47] the less accurate CDFT approach was used. From Tab. II one can see that according to DFT/PBE the rs phase is the most stable phase for all alkali halides. However, in reality CsCl, CsBr, and CsI crystallize in the cscl phase under the ambient conditions. In Ref. [48] it was pointed out that the dispersion correction can cure this problem. In the Appendix we also provide results obtained with the dispersion correction (see Tab. IV) which confirm this finding: The DFT/PBE+D2 $\Delta E_{\rm rs-cscl}$ has the correct sign (plus) for CsCl, CsBr, and CsI, as well as for all the others alkali halides (minus). We have relied on DFT/PBE+D2 results for locating rs to cscl transition composition of $AX$ binary and ternary mixtures in Secs. IV D, IV E.

A direct comparison of energy difference, $\Delta E_{\rm rs-cscl}$, to experimental data is obstructed, since a particular $AX$ occurs in the rs and cscl phase at a different pressure. However, we can qualitatively check the correlation between the experimental transition pressure and the calculated $\Delta E_{\rm rs-cscl}$. Generally, rs to cscl phase transition occurs at a lower pressure for $AX$ which consist of heavier elements[54, 71–73]. This is in agreement with our DFT/PBE results when looking at the columns of Tab. II—apart from alkali fluorides. The same trend is observed within the rows of Tab. II for all $AX$, but for Na$X$ and KI. However, the results for sodium halides are not inconsistent with experimental observation: NaBr and NaI transform to TlI[74] structure, and the experimental transition pressure for NaF is lower than for NaCl (23[75] vs. 27 GPa[76]).

Most of the calculated bulk moduli, reported in Tab. III, underestimate the experimental values, apart from K$X$ and Rb$X$ in cscl phase. However, all trends are caught by the DFT calculations: I.e. the bulk modulus systematically decreases when going from light to heavy atoms.

In summary, DFT/PBE (or DFT/PBE+D2) yields reasonable results for alkali halides when compared to experiment. Since we investigate the predictive power of alchemical first order derivatives for reproducing results coming from pure reference calculations, rather than reproducing experimental outcomes, we consider the DFT level of accuracy sufficiently accurate for this purpose.

TABLE IV. Alchemical predictions of $\Delta E_{\text{rs}-\text{cscl}}$/atom [meV] for $AX$. Columns indicate target $AX$ and rows correspond to the reference $AX$. The values on the diagonal are pure DFT/PBE+D2 calculations. MAPE is the percentage representation of MAE, in MAPE[1] the contributions from $AF$ were eliminated, in MAPE[2] the contributions from $AF$ and Na$X$ were eliminated, and in MAPE[3] the contributions from Cs$X$.

| | NaF | NaCl | NaBr | NaI | KF | KCl | KBr | KI | RbF | RbCl | RbBr | RbI | CsF | CsCl | CsBr | CsI | MAE | MAPE |
|---|---|---|---|---|---|---|---|---|---|---|---|---|---|---|---|---|---|---|
| NaF | -143 | -42 | -168 | -327 | -64 | 9 | -157 | -271 | -58 | 18 | -115 | -202 | -90 | 1 | -138 | -188 | 86 | 358.1 |
| NaCl | -121 | -172 | -163 | -122 | -50 | -62 | -66 | -61 | -15 | -34 | -35 | -22 | 6 | -2 | -17 | 39 | 27 | 76.5 |
| NaBr | -115 | -158 | -182 | -176 | -26 | -46 | -71 | -93 | -13 | -32 | -62 | -42 | -21 | -20 | -4 | 19 | 25 | 69.2 |
| NaI | -99 | -126 | -156 | -212 | -10 | -22 | -50 | -87 | 10 | 0 | -3 | -34 | 28 | 22 | 13 | -28 | 34 | 81.9 |
| KF | -112 | -27 | 80 | 196 | -65 | -20 | 75 | 189 | -54 | -7 | 90 | 199 | -50 | 25 | 115 | 228 | 135 | 396.4 |
| KCl | -187 | -120 | -92 | -45 | -49 | -59 | -54 | -32 | -20 | -29 | -30 | -13 | -7 | -12 | -7 | 12 | 40 | 61.9 |
| KBr | -158 | -138 | -126 | -85 | -49 | -55 | -58 | -57 | -14 | -23 | -32 | -31 | -3 | -12 | -10 | -8 | 31 | 63.6 |
| KI | -147 | -147 | -149 | -134 | -38 | -41 | -51 | -73 | -1 | -7 | -16 | -37 | 2 | -5 | -9 | -17 | 29 | 73.3 |
| RbF | -75 | -33 | 8 | 52 | -57 | -25 | 9 | 50 | -58 | -12 | 17 | 53 | -58 | 12 | 38 | 70 | 77 | 144.7 |
| RbCl | -259 | -108 | -69 | -28 | -107 | -60 | -47 | -23 | -32 | -24 | -24 | -7 | 7 | 7 | 5 | 15 | 50 | 57.3 |
| RbBr | -231 | -149 | -107 | -57 | -103 | -73 | -62 | -42 | -21 | -24 | -28 | -22 | 19 | 8 | 4 | 4 | 40 | 46.5 |
| RbI | -211 | -196 | -168 | -112 | -101 | -83 | -76 | -73 | -15 | -16 | -18 | -35 | 25 | 21 | 13 | -1 | 31 | 47.0 |
| CsF | -30 | -19 | -6 | 31 | -6 | -16 | 0 | 28 | -36 | -8 | 6 | 31 | -82 | 16 | 19 | 43 | 75 | 114.2 |
| CsCl | -223 | -77 | -39 | -10 | -180 | -58 | -31 | -8 | -102 | -31 | -18 | -2 | -12 | 12 | 14 | 19 | 60 | 71.0 |
| CsBr | -226 | -133 | -75 | -27 | -188 | -103 | -61 | -26 | -101 | -53 | -30 | -13 | 6 | 15 | 13 | 12 | 55 | 67.2 |
| CsI | -237 | -203 | -157 | -75 | -205 | -170 | -128 | -67 | -107 | -88 | -65 | -36 | 14 | 18 | 16 | 8 | 58 | 91.4 |
| MAE | 60 | 68 | 90 | 166 | 49 | 30 | 35 | 66 | 38 | 17 | 29 | 47 | 74 | 12 | 28 | 45 | | |
| MAPE | 41.9 | 39.3 | 49.2 | 78.3 | 74.9 | 51.3 | 60.9 | 90.0 | 64.7 | 72.2 | 102.6 | 133.5 | 90.4 | 73.3 | 213.8 | 559.2 | | |
| MAPE[1] | 46.9 | 39.9 | 58.8 | 88.4 | 86.7 | 51.3 | 63.0 | 92.4 | 63.8 | 66.7 | 90.3 | 133.2 | 93.7 | 82.6 | 153.1 | 451.1 | | |
| MAPE[2] | | | 46.0 | 25.0 | 40.0 | 70.0 | 94.5 | 46.0 | 31.0 | 43.8 | 73.0 | 68.2 | 36.2 | 44.6 | 106.9 | 95.8 | 82.7 | 125.0 |
| MAPE[3] | 43.7 | 46.9 | 62.2 | 85.5 | 66.0 | 40.7 | 66.9 | 110.3 | 54.2 | 55.7 | 112.7 | 166.0 | 79.3 | 101.5 | 274.1 | 626.0 | | |

### B. Example: NaCl → KCl

An exemplary selection of vertical iso-valence-electronic alchemical prediction scans of energies is shown in Fig. 2 for all three combinations of rs, zb, and cscl. The bottom row corresponds to DFT/PBE calculations of the energy, as a function of the lattice constant for the reference salt NaCl: the middle row corresponds to DFT/PBE for the target salt KCl, and the top row corresponds to the alchemical prediction estimated with first order Hellmann–Feynman derivatives. Note the huge offset of the alchemically predicted curves in the top row, previously alluded to. The shape, positioning, and relative energy gap between different phases, Eq. 4, however, is very similar to the true target case (mid row). The agreement is particularly stunning for combinations which involve zb (mid and left-hand column): The reference curve has a strikingly different shape, yet its alchemical prediction is still in good agreement with the target curve.

We can quantify this agreement: The energy difference between rs and cscl phases for NaCl is -175 meV. Alchemical derivatives predict this difference to shrink to -89 meV in the case of KCl. The true DFT/PBE energy difference of KCl amounts to -84 meV. Similar predictive power is found for rs and zb, and cscl and zb. Also, the predicted values of the equilibrium lattice constants are in startling agreement with the DFT truth: They deviate at most by 0.07 Å. The bulk modulus calculated from $E^p$ matches well the target in case of rs phase: 1.636 vs. 1.608 [$10^{11} \frac{\text{dynes}}{\text{cm}^2}$], whereas for the reference system it is 2.290 [$10^{11} \frac{\text{dynes}}{\text{cm}^2}$]. The agreement is similar for zb (1.247 vs. 1.191 [$10^{11} \frac{\text{dynes}}{\text{cm}^2}$]), but worse for cscl (1.008 vs. 1.709 [$10^{11} \frac{\text{dynes}}{\text{cm}^2}$]). These remarkable findings have motivated us to perform a more comprehensive screen for all possible combinations of reference and target salts.

### C. All to all

To probe the predictive power of first order alchemical derivatives for crystal structures we have studied all possible transmutations between 16 alkali halides in rs, cscl, and zb structure. Fig. 3 features the total potential energies as a function of interatomic distance for all 16×15 combinations of reference/target salts, all reference salts, and all rs, zb, and cscl phases. As discussed in Sec. III lithium halides have been excluded from this screen. All in all, we examined 720 alchemical couplings. The results shown in Fig. 3 also contain a table reporting extracted properties (lattice constants, relative energies, and bulk moduli) for all alkali chlorides. Complete tables with these properties for all alkali halide transmutations can be found in Appendix, Tabs. V, VI, VII, VIII, IX, X, XI, XII, and XIII.

Specifically, we have scanned each crystal structure in rs, cscl, and zb phase electron density and total DFT

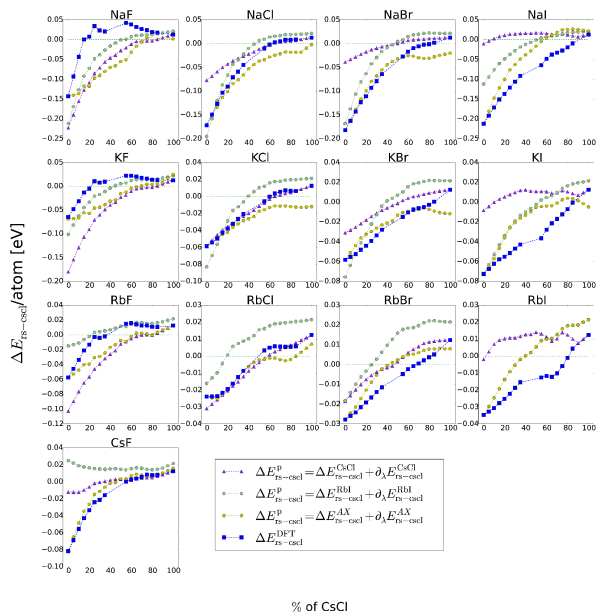

FIG. 5. Alchemical predictions and DFT/PBE+D2 results for $\Delta E_{\mathrm{rs-cscl}}$/atom [eV] as a function of binary mixture composition with $0 \leq x \leq 100\%$ for $(AX)_{1-x}(\mathrm{CsCl})_x$. Connecting lines are shown for convenience. Alchemical predictions, $\Delta E^{\mathrm{p}}_{\mathrm{rs-cscl}}$, are shown using the reference salts specified in the legend (CsCl, RbI, and $AX$).

energy for a wide range of lattice constants. This range has been chosen to run from a value, smaller than NaF equilibrium lattice constant to a value larger than CsI equilibrium lattice constant with a fixed spacing. The need to use such an extensive range arises from the fact that predictions from each salt are made for all other salts. Those points on the potential energy surface which lie too far away from equilibrium value are usually the most difficult to converge, and thus energy oscillations can occur. This naturally leads to oscillations in $E^p$, and therefore inaccurate results can occur for extreme combinations (see e.g. predictions of $A$Br or $A$Cl from NaF or KF in Fig. 3).

1. Relative energies

Alchemical predictions of relative energies are presented in Tabs. V, VI, and VII. These tables should be read as follows: The diagonal elements correspond to the true DFT/PBE result for the reference compound. All off-diagonal elements in any given row correspond to alchemical predictions obtained for the target salt specified in the head of the column, and using as a reference salt the diagonal element present in that same row. Overall we note that for the majority of combinations predictions are very accurate if the alkali ion is fixed, and only the halogen is allowed to vary. This corresponds to off-diagonal elements above or below the diagonal in multiples of four. For example, $\Delta E_{\mathrm{rs-cscl}}$ (Tabs. V) of NaBr (-181 meV/atom) is rather well predicted when using NaF (-182 meV/atom), NaCl (-162 meV/atom), or NaI (-160 meV/atom) as a reference. In other combinations, the important role of the choice of reference, however, becomes obvious. For example, if we exclude $A$F and Na$X$ from references, the percentage representation of MAE for predicting $\Delta E_{\mathrm{rs-cscl}}$ of CsI drops from 145.6 to 22.9 % when we exclude $A$F or Na$X$ from the reference salts.

Generally, the predictive accuracy is the worst if the target salt is composed out of heavy atoms (Cs$X$, Rb$X$) and the reference salt is composed out of light atoms without $d$ electrons ($A$F, Na$X$). This observation is consistent with recent findings in small molecules [38]. MAE due to choice of reference salt are reported in the out most right hand columns in the tables. The lowest MAE ($\sim$30 meV/atom) is found for reference salts RbI, NaBr, KBr, NaCl in the case of the rs-cscl energy difference. In the case of the rs-zb energy difference, the lowest MAE ($\sim$13 meV/atom) is found for reference salts RbF and CsF, followed by KCl, KBr, and NaCl ($\sim$23 meV/atom). For relative zb-cscl energies, the lowest MAE is obtained for NaCl (39 meV/atom) and NaBr (43 meV/atom) as a reference salt. To put these results into perspective we refer to the DFT analysis by Lany [77] who reported prediction errors for heats of formation for general chemistries with filled $d$-shells which (assuming normal distributions) amount to MAE of at least 0.19 eV/atom [78]. We note that for all the 16×15 alkali halide combinations in all the three combinations of energy differences (rs-cscl, rs-zb, and zb-cscl) no reference salt yields worse predictions than that, except for KF in the case of zb-cscl (MAE = 284 meV/atom). Similar DFT errors for solids were also reported by Mattsson and co-workers [79]. As such, our numerical evidence indicates that alchemical predictions of relative energies achieve a predictive power on par to generalized gradient based DFT (when compared to experiment).

In Tab. V one can note the aforementioned DFT/PBE artifact that the rs phase is preferable over the cscl for all alkali halides. Under normal conditions, of course, CsCl, CsBr, and CsI should favor the cscl phase. In Ref. [48] this shortcoming was described, and the authors pointed out that interatomic two-body $C_6/R^6$ dispersion corrections can cure this failure. We have reproduced this finding, and it is summarized in Tab. IV. When augmenting the alchemical predictions with D2 corrections, the predictive accuracy slightly increases: Reference salt NaBr has the lowest MAE of 25 meV/atom, NaCl is still a good reference with a decreased MAE of 27 meV/atom, and KBr and RbI are next with 31 meV/atom. We note though that NaBr, NaCl and KBr still fail to predict the correct sign for all the three salts which favor cscl structure, i.e. CsCl, CsBr, and CsI. RbI as a reference salt clearly accounts for CsCl and CsBr in the cscl, and it predicts practically zero energy difference for CsI. Since it represents the best compromise between yielding the correct sign and small overall MAE, we have therefore opted to probe RbI as the most promising reference for the binary and ternary mixtures (see below) for which the rs to cscl phase transition can be observed.

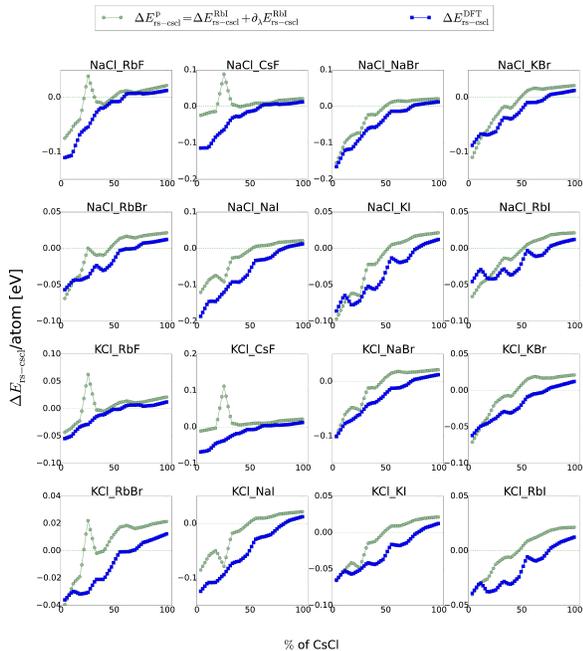

FIG. 6. Alchemical predictions and DFT/PBE+D2 results for $\Delta E_{\rm rs-cscl}$/atom [eV] are shown as a function of the $(AX)_{0.5-0.5x}(BY)_{0.5-0.5x}({\rm CsCl})_x$ ternary mixture composition with $0 \leqslant x \leqslant 100\%$. Top and bottom two rows correspond to $AX$ equals NaCl and KCl, respectively. All alchemical predictions result from RbI as a reference salt. Connecting lines are plotted for convenience. Outliers at 25% are artifacts.

### 2. Lattice constants

Equilibrium lattice constants were alchemically predicted by fitting the $E^{\rm p}$ curves in Fig. 3 and locating their minimum. Predictions are reported in Tabs. VIII, IX, X. The predictive accuracy of lattice constants in alkali halides can be remarkable, e.g. in the aforementioned example of NaCl→KCl. On average, however, predicted lattice constants are not of similar accuracy as DFT. Using CsF as a reference for estimating lattice constants of alkali halides in zb a MAE is obtained of 1Å—less than ten percent of this number would be desirable. The best reference on average, NaF, has a MAE less than 0.5 Å, which clearly still lacks quantitative accuracy. Qualitative trends (the heavier the elements the larger the lattice constant), however, are well reproduced. Inclusion of higher order effects might still improve the location of minima in the predicted energy curves, as it was shown to be the case for covalent bonds in small molecules [38]. However, we consider this shortcoming to be less severe: Decent structural information can often be obtained already at the level of interatomic potentials or semi-empirical methods such as tight-binding DFT.

### 3. Bulk moduli

The 720 predicted energy curves in Fig. 3 were fitted with BM equation of state in order to calculate bulk moduli for all predictions. Experimental and DFT data suggest that alkali halides' bulk modulus decreases as we go from lighter to heavier elements, i.e. fluorides to iodides for fixed alkali atom. Alchemical predictions of rs bulk moduli from all possible alkali halides for all possible Na$X$, K$X$, and Rb$X$ feature in Fig. 4. Predictions of Cs$X$ have been omitted since for them the cscl phase is more stable. Blue and red curves in Fig. 4 represent the experimental and DFT data. While there are some deviations in the predictions (up to $2.0 \times 10^{11} \frac{\rm dynes}{\rm cm^2}$), the overall trend is reproduced. Specific numerical results for all predictions of bulk moduli are shown in Tabs. XI, XII, XIII for rs, cscl, and zb structure, respectively. Interestingly, the reference salt CsCl yields the best predictive power with a MAE $< 0.4 \times 10^{11} \frac{\rm dynes}{\rm cm^2}$.

### D. Binary mixtures

When doping one alkali halide $AX$ with another alkali halide $BY$, one finds that relative DFT energies hardly depend on the spatial distribution of $BY$ in $AX$. This is not surprising due to the predominant ionic mode of binding in these crystals, containing only ions with the same charge (plus or minus one). Alchemical estimates have also confirmed this near degeneracy. Varying compositional degrees of freedom, however, leads to significant changes in relative energies. In particular, a phase transition rs → cscl occurs at some mole fraction $x_t$ in $(AX)_{1-x}({\rm CsCl})_x$. We have studied if alchemical estimates, based on DFT/PBE+D2 level of theory, can capture the rs → cscl phase transition as $x$ goes from zero to one, and if they could possibly even predict $x_t$. In our setup the rs and cscl crystal structures contain different number of atoms (64 vs. 54), and we have therefore encountered finite size effects: When substituting atoms in some $AX$ crystal structures with Cs and Cl atoms, the exact percentage of CsCl will differ for the rs and cscl phase. In order to obtain energy differences for the same component ratio $x$ we have linearly interpolated the energy curves, and averaged the result for rs and cscl phase. $\Delta E_{\rm rs-cscl}$ is shown in Fig. 5 as a function of CsCl percentage for all binary mixtures with alkali halides which favor rs. The component ratio is varying from 0 to 100% with a step of 5%.

We compare DFT results with alchemical predictions of $\Delta E_{\rm rs-cscl}$ in $(AX)_{1-x}({\rm CsCl})_x$ made using RbI, CsCl, and $AX$ as reference salts in Fig. 5. One can observe a remarkable overall correlation between DFT and alchemical predictions.

Except for $CsF_{1-x}CsCl_x$, RbI always predicts the correct change of sign. CsF is the odd-one-out among Cs halides: It favors rs. In Sec. IV C we analyzed the overall performance of energy differences for pure $AX$ and noticed that alkali fluorides correspond to the most demanding task for alchemy, especially CsF. We believe that the ultimate reason for this is the lack of $d$-electrons which makes the fluoride ion different from all the other halogens. As such, it represents a special case for RbI since RbI alchemically predicts it to be in cscl structure already in its pure state (See Tab. IV). While RbI captures the correct qualitative trend of phase stability for all the typical rs alkali halides it does predict transition ratios $x_t$ with a systematic shift. In particular, predicted $x_t$-values are typically ∼0.4 too small with respect to DFT's $x_t$-value when estimating relative energies for $(AX)_{1-x}(CsCl)_x$ where $X \in \{Cl, Br, I\}$. When predicting $x_t$ in $(AF)_{1-x}(CsCl)_x$, RbI typically overshoots with respect to DFT.

Similar observations can be made for CsCl reference: When predicting $x_t$ in $(AF)_{1-x}(CsCl)_x$, CsCl typically overshoots with respect to DFT. For $(ACl)_{1-x}(CsCl)_x$, $x_t$ of DFT is very well reproduced by alchemical estimates based on CsCl. And for $(ABr)_{1-x}(CsCl)_x$ and $(AI)_{1-x}(CsCl)_x$, the CsCl based estimate of $x_t$ is systematically underestimated, i.e. for I more severely so than for Br.

When using $AX$ as a reference salt, the overall agreement of alchemically predicted relative energy curves with DFT results is not poor. The predictions lacks severely, however, in predictive power when it comes to the sign change or to $x_t$. Only $CsF_{1-x}CsCl_x$ is predicted very well using CsF as a reference. This is consistent with the fact, on display in Tab. IV, that only $AF$ reference salts yield satisfactory predictions of CsF, while the majority of other reference salts predict the wrong sign. With increasing CsCl component the prediction becomes better and can hardly be distinguished from DFT results after reaching 50% CsCl fraction.

In summary, we believe that these results amount to numerical evidence which suggests that it is possible to predict if phase transitions will occur when using a *single* and *pure* reference salt to screen the entire binary mixture space. RbI, the $AX$ with the heaviest elements, turned out to be a good reference salt choice for alkali halides. It remains to be seen, if this observation also holds for other crystals and mixtures.

### E. Ternary mixtures

In order to explore the limits of our approach, we have extended the binary to the ternary search space. Obviously, the larger the number of components in a multi-component mixture, the more efficient a predictive alchemical screening tool which is based on a single pure reference salt. In order to keep the reference DFT calculations tractable, we had to severely restrict ourselves in the ternary compound space. More specifically, we have considered the admixture of CsCl with a fixed 50-50% ratio of $AX$, $BY$. For the alchemical screening of $\Delta E_{\text{rs-cscl}}$ as a function of CsCl content $x$, we have selected sixteen ternary mixtures containing NaCl and KCl, respectively, i.e. eight mixtures $NaCl_{0.5-0.5x}AX_{0.5-0.5x}(CsCl)_x$ and eight mixtures $KCl_{0.5-0.5x}AX_{0.5-0.5x}(CsCl)_x$. $\Delta E_{\text{rs-cscl}}$ as a function of CsCl content $x$ is shown in Fig. 6 for DFT/PBE+D2 and alchemical predictions based on pure RbI as a reference salt.

In striking similarity to the binary case, overall correlation of DFT curves with alchemical predictions is striking. Alchemical predictions based on RbI reference reproduce the phase transition for all mixtures shown. The alchemical relative energies systematically overshoot the DFT results, resulting into an estimated $x_t$ which is systematically too small (on average $\sim 0.4$). We note an outlier in the alchemical predictions at 25% CsCl content for all curves shown. This is due to a finite size effect of the differing numbers of atoms/unit cell. For our unit cells this effect becomes sizable already for ternary mixtures. We expect this effect to disappear if larger unit cells were chosen. Finally, we reiterate that all the alchemical prediction curves in Fig. 6 have been obtained from alchemical predictions based on lattice scans (rs and cscl) of just a single reference system!

## V. CONCLUSION

We have studied the predictive power of first order Hellmann–Feynman based "alchemical" derivatives for iso-valence-electronic changes in alkali halide crystal structures. Specifically, we examined properties, such as relative energies, lattice parameters, and bulk moduli for rs, cscl, and zb phases. For relative energies between rs and cscl phases, we also studied binary and ternary mixtures with CsCl.

We find that vertical alchemical predictions of relative energies reach an accuracy which is on par, if not better, than what one can expect from DFT, ∼ 0.1 eV/atom. We presume that this accuracy can be reached by first order derivatives due to the cancellation of non-linear effects, present in the electron density response, when considering relative energies. Similar to earlier studies, Ref. [36, 38], we observed the best accuracy when inter converting elements that appear late in the periodic table. The choice of the reference compound (whose electron density is used to make the prediction) is crucial for the predictive power of alchemical estimates. Reference compounds from late row elements typically result in accurate predictions, on average, however, NaCl is best. Predicted lattice parameters deviate substantially (MAE of the order of 0.5 Å for the best reference compound, NaF). For bulk moduli, however, the MAE obtained for the best reference compound CsCl is quite acceptable. The reversal of the rs-cscl phase stability trend when admixing CsCl to alkali halides is captured by alchemical predictions when using RbI as a single and pure reference salt. Note that in order to account for this effect already within DFT, van der Waals corrected DFT has been necessary, as also pointed out previously by others [48]. Alchemical predictions based on RbI reference also capture the phase transition in ternary mixtures with CsCl.

Overall, our results suggest that the pragmatic use of alchemical couplings holds great promise for very efficient screening campaigns which explore the materials compound space spanned by multi-component ionic crystals. It remains to be seen if alchemical predictions can also be applied in the context of other solids, or even liquids.

## VI. ACKNOWLEDGMENTS

The authors thank A. Michaelides for suggesting to include the rs/cscl phases in our investigation. OAvL acknowledges funding from the Swiss National Science foundation (No. PP00P2_138932). Some calculations were performed at sciCORE (http://scicore.unibas.ch/) scientific computing core facility at University of Basel and at Swiss National Supercomputing Centre (http://www.cscs.ch/).

Appendix: Numerical results

In the following, we provide all results for true and predicted (with first order alchemical derivatives) energy differences (Tabs. V, VI, VII, IV), lattice constants (Tabs. VIII, IX, X), and bulk moduli (Tabs. XI, XII, XIII) for pure alkali halides in rs, cscl, and zb phases. Mean absolute errors are given in each table for each reference and target alkali halide. We also provide mean absolute percentage error (MAPE), which is the deviation of the forecast value from the DFT benchmark in percentage.

TABLE V. Alchemical predictions of $\Delta E_{\rm rs-cscl}$/atom [meV] for $AX$. Columns indicate target $AX$ and rows correspond to the reference $AX$. The values on the diagonal are pure DFT/PBE calculations. MAPE is the percentage representation of MAE, in MAPE[1] the contributions from $AF$ were eliminated, and in MAPE[2] the contributions from $AF$ and Na$X$ were eliminated.

|      | NaF  | KF   | RbF  | CsF  | NaCl | KCl  | RbCl | CsCl | NaBr | KBr  | RbBr | CsBr | NaI  | KI   | RbI  | CsI  | MAE | MAPE |
|------|------|------|------|------|------|------|------|------|------|------|------|------|------|------|------|------|-----|------|
| NaF  | -163 | -99  | -87  | -71  | -46  | -26  | -31  | -68  | -182 | -192 | -171 | -212 | -313 | -300 | -245 | -262 | 96  | 168.2 |
| KF   | -124 | -100 | -90  | -83  | -13  | -35  | -36  | -29  | 79   | 71   | 72   | 58   | 185  | 176  | 174  | 160  | 134 | 174.4 |
| RbF  | -78  | -89  | -94  | -92  | -36  | -37  | -38  | -30  | 6    | 6    | 5    | -2   | 43   | 43   | 43   | 41   | 83  | 91.3 |
| CsF  | -29  | -39  | -72  | -106 | -21  | -22  | -23  | -25  | -7   | -7   | -7   | -7   | 23   | 23   | 23   | 24   | 85  | 89.4 |
| NaCl | -123 | -59  | -43  | -38  | -175 | -89  | -74  | -66  | -162 | -97  | -80  | -91  | -110 | -83  | -70  | -37  | 31  | 39.6 |
| KCl  | -175 | -73  | -47  | -46  | -119 | -84  | -64  | -53  | -94  | -75  | -60  | -50  | -52  | -44  | -38  | -32  | 36  | 31.7 |
| RbCl | -248 | -127 | -60  | -38  | -103 | -81  | -60  | -46  | -73  | -61  | -52  | -42  | -34  | -33  | -31  | -26  | 45  | 37.1 |
| CsCl | -215 | -194 | -129 | -56  | -77  | -70  | -61  | -40  | -46  | -45  | -40  | -35  | -22  | -22  | -23  | -20  | 55  | 46.3 |
| NaBr | -116 | -30  | -38  | -38  | -158 | -69  | -70  | -56  | -181 | -96  | -86  | -57  | -162 | -112 | -78  | -45  | 30  | 37.3 |
| KBr  | -151 | -58  | -35  | -33  | -136 | -76  | -50  | -43  | -120 | -82  | -58  | -48  | -83  | -70  | -55  | -43  | 31  | 28.6 |
| RbBr | -220 | -109 | -48  | -6   | -146 | -95  | -57  | -34  | -107 | -82  | -58  | -40  | -62  | -55  | -46  | -36  | 35  | 29.0 |
| CsBr | -217 | -189 | -115 | -36  | -126 | -115 | -82  | -39  | -77  | -73  | -57  | -36  | -34  | -34  | -33  | -25  | 46  | 39.2 |
| NaI  | -98  | -26  | 3    | -13  | -124 | -42  | -8   | -22  | -160 | -64  | -25  | -38  | -194 | -101 | -71  | -61  | 42  | 50.2 |
| KI   | -140 | -45  | -17  | -28  | -140 | -52  | -26  | -34  | -137 | -70  | -36  | -37  | -128 | -89  | -58  | -40  | 33  | 33.5 |
| RbI  | -205 | -102 | -32  | -9   | -184 | -98  | -40  | -15  | -156 | -95  | -46  | -21  | -107 | -84  | -57  | -35  | 29  | 30.8 |
| CsI  | -232 | -194 | -115 | -21  | -196 | -171 | -103 | -27  | -151 | -132 | -87  | -27  | -79  | -72  | -58  | -30  | 46  | 46.1 |
| MAE  | 54   | 46   | 43   | 65   | 71   | 32   | 22   | 13   | 89   | 41   | 35   | 32   | 148  | 71   | 51   | 44   |     |      |
| MAPE | 33.4 | 46.5 | 45.3 | 61.8 | 40.4 | 37.9 | 36.1 | 32.8 | 49.0 | 50.6 | 60.9 | 87.6 | 76.0 | 69.9 | 89.5 | 145.6 |    |      |
| MAPE[1] | 28.5 | 52.0 | 53.7 | 71.5 | 24.7 | 28.4 | 32.3 | 30.2 | 35.6 | 21.7 | 27.1 | 35.4 | 59.1 | 34.6 | 26.5 | 32.7 |  |      |
| MAPE[2] | 27.7 | 48.8 | 47.5 | 71.4 | 25.9 | 29.8 | 28.5 | 22.8 | 41.0 | 22.7 | 19.4 | 21.5 | 65.6 | 41.9 | 25.9 | 22.9 |  |      |

TABLE VI. Alchemical predictions of $\Delta E_{\rm rs-zb}$/atom [meV] for $AX$. Columns indicate target $AX$ and rows correspond to the reference $AX$. The values on the diagonal are pure DFT/PBE calculations. MAPE is the percentage representation of MAE, in MAPE[1] the contributions from $AF$ were eliminated, and in MAPE[2] the contributions from $AF$ and Na$X$ were eliminated.

|      | NaF  | KF   | RbF  | CsF  | NaCl | KCl  | RbCl | CsCl | NaBr | KBr  | RbBr | CsBr | NaI  | KI   | RbI  | CsI  | MAE | MAPE |
|------|------|------|------|------|------|------|------|------|------|------|------|------|------|------|------|------|-----|------|
| NaF  | -83  | -93  | -117 | -108 | -84  | -76  | -79  | -120 | -290 | -300 | -263 | -302 | -478 | -468 | -426 | -465 | 173 | 255.2 |
| KF   | -125 | -81  | -79  | -93  | -91  | -87  | -86  | -82  | -275 | -253 | -243 | -237 | -450 | -430 | -420 | -409 | 150 | 222.5 |
| RbF  | -130 | -87  | -69  | -60  | -84  | -83  | -82  | -77  | -84  | -84  | -85  | -86  | -82  | -82  | -82  | -84  | 13  | 22.2 |
| CsF  | -120 | -109 | -80  | -35  | -85  | -84  | -81  | -75  | -82  | -81  | -80  | -77  | -85  | -85  | -84  | -83  | 13  | 19.6 |
| NaCl | -86  | -114 | -129 | -112 | -67  | -84  | -122 | -110 | -63  | -91  | -102 | -116 | -56  | -74  | -96  | -89  | 24  | 39.2 |
| KCl  | -135 | -85  | -90  | -107 | -101 | -92  | -96  | -106 | -94  | -90  | -92  | -104 | -84  | -83  | -87  | -96  | 23  | 41.5 |
| RbCl | -239 | -121 | -76  | -81  | -112 | -96  | -86  | -89  | -95  | -88  | -86  | -88  | -84  | -83  | -85  | -88  | 27  | 43.1 |
| CsCl | -225 | -197 | -131 | -63  | -117 | -110 | -98  | -74  | -95  | -93  | -88  | -79  | -85  | -84  | -82  | -81  | 35  | 50.8 |
| NaBr | -89  | -112 | -123 | -132 | -70  | -103 | -116 | -126 | -59  | -95  | -119 | -108 | -51  | -83  | -97  | -72  | 26  | 43.2 |
| KBr  | -117 | -87  | -95  | -106 | -106 | -91  | -93  | -105 | -93  | -87  | -92  | -106 | -85  | -86  | -89  | -98  | 23  | 40.9 |
| RbBr | -195 | -106 | -82  | -66  | -141 | -105 | -89  | -88  | -107 | -95  | -87  | -89  | -88  | -87  | -85  | -89  | 27  | 42.2 |
| CsBr | -217 | -190 | -123 | -65  | -151 | -141 | -111 | -78  | -112 | -108 | -96  | -80  | -88  | -87  | -87  | -80  | 41  | 60.1 |
| NaI  | -87  | -113 | -115 | -114 | -76  | -107 | -111 | -111 | -67  | -107 | -105 | -109 | -44  | -89  | -99  | -93  | 23  | 38.3 |
| KI   | -99  | -92  | -97  | -116 | -98  | -89  | -96  | -115 | -89  | -90  | -94  | -113 | -83  | -86  | -91  | -100 | 24  | 42.9 |
| RbI  | -175 | -111 | -87  | -91  | -162 | -111 | -88  | -91  | -133 | -102 | -88  | -89  | -96  | -90  | -85  | -89  | 33  | 54.0 |
| CsI  | -225 | -195 | -132 | -74  | -200 | -179 | -127 | -79  | -160 | -145 | -113 | -79  | -105 | -101 | -95  | -81  | 60  | 86.6 |
| MAE  | 68   | 40   | 35   | 58   | 45   | 18   | 14   | 23   | 64   | 36   | 31   | 39   | 89   | 52   | 50   | 55   |     |      |
| MAPE | 81.8 | 49.1 | 50.3 | 164.4 | 67.0 | 19.3 | 16.8 | 30.8 | 107.8 | 41.0 | 35.3 | 49.3 | 203.0 | 60.3 | 58.4 | 67.6 |   |      |
| MAPE[1] | 89.7 | 56.7 | 54.6 | 168.3 | 81.0 | 22.5 | 21.2 | 34.9 | 70.7 | 15.4 | 12.5 | 23.2 | 87.0 | 5.0 | 6.8 | 11.7 |   |      |
| MAPE[2] | 117.8 | 62.4 | 47.0 | 144.1 | 97.0 | 26.4 | 16.0 | 26.9 | 84.2 | 16.5 | 7.9 | 17.3 | 101.5 | 4.2 | 4.0 | 11.6 |   |      |

TABLE VII. Alchemical predictions of $\Delta E_{\text{zb-cscl}}$/atom [meV] for $AX$. Columns indicate target $AX$ and rows correspond to the reference $AX$. The values on the diagonal are pure DFT/PBE calculations. MAPE is the percentage representation of MAE, in MAPE[1] the contributions from $AF$ were eliminated, and in MAPE[2] the contributions from $AF$ and Na$X$ were eliminated.

| | NaF | KF | RbF | CsF | NaCl | KCl | RbCl | CsCl | NaBr | KBr | RbBr | CsBr | NaI | KI | RbI | CsI | MAE | MAPE |
|---|---|---|---|---|---|---|---|---|---|---|---|---|---|---|---|---|---|---|
| NaF | -79 | -6 | 30 | 37 | 38 | 50 | 48 | 52 | 108 | 108 | 92 | 89 | 165 | 169 | 181 | 203 | 109 | 706.0 |
| KF | 1 | -19 | -12 | 10 | 78 | 52 | 50 | 54 | 354 | 323 | 315 | 295 | 634 | 606 | 594 | 569 | 284 | 2220.2 |
| RbF | 52 | -2 | -25 | -32 | 49 | 47 | 44 | 47 | 91 | 90 | 90 | 85 | 125 | 125 | 125 | 125 | 92 | 544.4 |
| CsF | 91 | 70 | 8 | -71 | 64 | 62 | 58 | 50 | 74 | 74 | 73 | 70 | 108 | 107 | 107 | 107 | 94 | 519.9 |
| NaCl | -36 | 56 | 86 | 74 | -108 | -5 | 48 | 44 | -99 | -5 | 22 | 25 | -54 | -9 | 27 | 51 | 39 | 128.3 |
| KCl | -40 | 12 | 43 | 61 | -18 | 8 | 32 | 53 | 0 | 15 | 32 | 54 | 32 | 39 | 49 | 63 | 52 | 185.4 |
| RbCl | -9 | -7 | 16 | 42 | 9 | 15 | 26 | 43 | 22 | 27 | 34 | 46 | 49 | 50 | 55 | 61 | 55 | 219.3 |
| CsCl | 10 | 3 | 2 | 7 | 40 | 39 | 36 | 35 | 50 | 48 | 48 | 44 | 63 | 62 | 59 | 61 | 64 | 301.0 |
| NaBr | -27 | 81 | 85 | 93 | -88 | 34 | 46 | 70 | -122 | -1 | 32 | 50 | -111 | -29 | 19 | 26 | 43 | 193.4 |
| KBr | -34 | 29 | 60 | 73 | -31 | 15 | 43 | 62 | -27 | 5 | 34 | 58 | 2 | 15 | 34 | 56 | 50 | 134.2 |
| RbBr | -26 | -3 | 34 | 60 | -5 | 11 | 32 | 54 | 1 | 13 | 29 | 49 | 27 | 32 | 39 | 53 | 50 | 158.8 |
| CsBr | 0 | 1 | 7 | 29 | 25 | 25 | 29 | 38 | 35 | 35 | 38 | 44 | 53 | 53 | 54 | 55 | 58 | 246.2 |
| NaI | -11 | 87 | 118 | 102 | -48 | 65 | 103 | 89 | -92 | 42 | 80 | 71 | -150 | -12 | 28 | 32 | 61 | 267.8 |
| KI | -41 | 48 | 80 | 87 | -42 | 37 | 70 | 81 | -48 | 20 | 58 | 76 | -45 | -3 | 32 | 61 | 55 | 160.3 |
| RbI | -30 | 9 | 55 | 81 | -22 | 13 | 48 | 76 | -23 | 7 | 42 | 68 | -11 | 6 | 28 | 54 | 50 | 113.8 |
| CsI | -7 | 0 | 17 | 53 | 4 | 8 | 23 | 52 | 9 | 13 | 26 | 51 | 27 | 29 | 36 | 51 | 50 | 147.0 |
| MAE | 72 | 44 | 67 | 123 | 112 | 25 | 22 | 23 | 152 | 51 | 40 | 34 | 221 | 91 | 69 | 60 | | |
| MAPE | 91.0 | 232.6 | 267.7 | 173.0 | 103.3 | 311.7 | 83.6 | 64.8 | 124.9 | 1021.3 | 138.2 | 77.1 | 147.3 | 3044.4 | 247.4 | 117.6 | | |
| MAPE[1] | 73.5 | 238.6 | 301.0 | 189.4 | 85.2 | 221.6 | 80.4 | 71.9 | 87.2 | 347.3 | 46.1 | 30.2 | 101.9 | 1063.6 | 46.8 | 17.8 | | |
| MAPE[2] | 75.1 | 153.8 | 239.6 | 177.2 | 95.9 | 154.7 | 53.4 | 63.9 | 101.7 | 345.0 | 28.3 | 26.7 | 114.6 | 1291.7 | 59.8 | 13.7 | | |

TABLE VIII. Alchemical predictions of alkali halides' lattice constants [Å] in rs symmetry. Columns indicate target $AX$ and rows correspond the reference $AX$. The values on the diagonal are pure DFT/PBE calculations.

| | NaF | KF | RbF | CsF | NaCl | KCl | RbCl | CsCl | NaBr | KBr | RbBr | CsBr | NaI | KI | RbI | CsI | MAE | MAPE |
|---|---|---|---|---|---|---|---|---|---|---|---|---|---|---|---|---|---|---|
| NaF | 4.80 | 5.40 | 5.60 | 5.80 | 6.00 | 6.20 | 6.40 | 6.60 | 6.60 | 6.60 | 6.60 | 6.60 | 6.60 | 6.60 | 6.60 | 6.60 | 0.45 | 6.4 |
| KF | 5.20 | 5.40 | 5.80 | 6.20 | 6.60 | 6.80 | 6.80 | 7.00 | 7.00 | 7.00 | 7.00 | 7.40 | 7.40 | 7.40 | 7.40 | 8.00 | 0.29 | 4.7 |
| RbF | 5.40 | 5.40 | 5.80 | 6.20 | 7.00 | 7.00 | 7.00 | 7.00 | 7.40 | 7.60 | 7.60 | 7.60 | 8.00 | 8.00 | 8.00 | 8.20 | 0.58 | 9.1 |
| CsF | 5.60 | 5.80 | 5.80 | 6.20 | 7.20 | 7.20 | 7.20 | 7.40 | 7.80 | 7.80 | 7.80 | 7.80 | 8.40 | 8.40 | 8.40 | 8.40 | 0.85 | 13.2 |
| NaCl | 5.30 | 6.10 | 6.60 | 7.00 | 5.70 | 6.40 | 6.70 | 7.00 | 6.20 | 6.60 | 6.70 | 7.00 | 6.60 | 7.00 | 7.00 | 7.00 | 0.37 | 5.9 |
| KCl | 5.60 | 6.00 | 6.60 | 7.20 | 6.20 | 6.40 | 6.80 | 7.20 | 6.60 | 6.80 | 7.00 | 7.40 | 7.20 | 7.40 | 7.40 | 7.80 | 0.37 | 6.3 |
| RbCl | 5.40 | 5.80 | 6.40 | 7.00 | 6.20 | 6.40 | 6.80 | 7.20 | 6.80 | 7.00 | 7.40 | 7.60 | 7.60 | 7.60 | 7.60 | 7.80 | 0.37 | 6.1 |
| CsCl | 5.80 | 6.00 | 6.20 | 6.60 | 6.60 | 6.60 | 6.80 | 7.00 | 7.20 | 7.20 | 7.20 | 7.40 | 7.80 | 7.80 | 8.00 | 8.00 | 0.51 | 8.5 |
| NaBr | 5.60 | 6.40 | 7.00 | 7.00 | 5.80 | 6.60 | 7.00 | 7.00 | 6.00 | 6.60 | 7.00 | 7.00 | 6.60 | 7.00 | 7.00 | 7.60 | 0.39 | 6.5 |
| KBr | 5.80 | 6.20 | 6.80 | 7.40 | 6.20 | 6.40 | 7.00 | 7.60 | 6.40 | 6.80 | 7.00 | 7.60 | 7.20 | 7.20 | 7.40 | 7.80 | 0.45 | 7.6 |
| RbBr | 5.80 | 6.00 | 6.60 | 7.00 | 6.20 | 6.40 | 6.80 | 7.40 | 6.60 | 6.80 | 7.00 | 7.40 | 7.40 | 7.40 | 7.60 | 7.80 | 0.38 | 6.5 |
| CsBr | 6.00 | 6.20 | 6.40 | 7.00 | 6.60 | 6.60 | 6.80 | 7.20 | 7.00 | 7.00 | 7.20 | 7.40 | 7.80 | 7.80 | 7.80 | 8.00 | 0.55 | 9.3 |
| NaI | 6.00 | 7.00 | 7.20 | 7.60 | 6.00 | 7.00 | 7.20 | 7.60 | 6.20 | 7.00 | 7.20 | 7.60 | 6.60 | 7.00 | 7.60 | 7.60 | 0.58 | 9.8 |
| KI | 6.20 | 6.60 | 7.20 | 8.40 | 6.40 | 6.60 | 7.20 | 8.40 | 6.60 | 6.80 | 7.40 | 8.40 | 7.00 | 7.20 | 7.60 | 8.20 | 0.78 | 12.8 |
| RbI | 6.20 | 6.40 | 7.00 | 7.60 | 6.20 | 6.60 | 7.00 | 7.60 | 6.40 | 6.80 | 7.20 | 7.80 | 7.20 | 7.20 | 7.60 | 8.00 | 0.55 | 9.4 |
| CsI | 6.40 | 6.40 | 6.80 | 7.40 | 6.60 | 6.60 | 6.80 | 7.40 | 6.80 | 6.80 | 7.00 | 7.40 | 7.40 | 7.40 | 7.60 | 7.80 | 0.54 | 9.4 |
| MAE | 0.95 | 0.71 | 0.76 | 0.88 | 0.69 | 0.25 | 0.17 | 0.36 | 0.77 | 0.24 | 0.22 | 0.27 | 0.75 | 0.37 | 0.31 | 0.31 | | |
| MAPE | 19.9 | 13.2 | 13.1 | 14.2 | 12.0 | 4.0 | 2.5 | 5.1 | 12.9 | 3.5 | 3.1 | 3.6 | 11.3 | 5.2 | 4.0 | 3.9 | | |

TABLE IX. Alchemical predictions of alkali halides' lattice constants [Å] in cscl symmetry. Columns indicate target $AX$ and rows correspond to the reference $AX$. The values on the diagonal are pure DFT/PBE calculations.

|      | NaF  | KF   | RbF  | CsF  | NaCl | KCl  | RbCl | CsCl | NaBr | KBr  | RbBr | CsBr | NaI  | KI   | RbI  | CsI  | MAE  | MAPE |
|------|------|------|------|------|------|------|------|------|------|------|------|------|------|------|------|------|------|------|
| NaF  | 2.93 | 3.27 | 3.40 | 3.47 | 3.60 | 3.67 | 3.80 | 3.87 | 3.87 | 3.87 | 3.87 | 4.13 | 4.13 | 4.40 | 4.40 | 4.40 | 0.16 | 3.9  |
| KF   | 3.07 | 3.27 | 3.47 | 3.73 | 4.20 | 4.00 | 4.20 | 4.27 | 4.27 | 4.27 | 4.27 | 4.40 | 4.47 | 4.47 | 4.47 | 4.47 | 0.17 | 4.5  |
| RbF  | 3.20 | 3.33 | 3.47 | 3.73 | 4.13 | 4.13 | 4.20 | 4.27 | 4.47 | 4.47 | 4.47 | 4.53 | 4.80 | 4.80 | 4.80 | 4.87 | 0.30 | 7.6  |
| CsF  | 3.40 | 3.40 | 3.53 | 3.73 | 4.27 | 4.27 | 4.33 | 4.33 | 4.67 | 4.60 | 4.67 | 4.67 | 5.07 | 5.07 | 5.07 | 5.07 | 0.46 | 11.7 |
| NaCl | 3.20 | 3.73 | 3.93 | 4.07 | 3.60 | 3.80 | 3.93 | 4.07 | 3.80 | 3.93 | 4.07 | 4.07 | 4.13 | 4.27 | 4.27 | 4.40 | 0.21 | 5.5  |
| KCl  | 3.47 | 3.60 | 3.87 | 4.27 | 3.73 | 3.87 | 4.07 | 4.33 | 4.00 | 4.07 | 4.20 | 4.47 | 4.40 | 4.40 | 4.47 | 4.67 | 0.19 | 5.3  |
| RbCl | 3.40 | 3.53 | 3.73 | 4.13 | 3.80 | 3.87 | 4.00 | 4.27 | 4.13 | 4.13 | 4.20 | 4.40 | 4.47 | 4.53 | 4.53 | 4.67 | 0.17 | 4.9  |
| CsCl | 3.60 | 3.73 | 3.73 | 4.00 | 4.00 | 4.00 | 4.07 | 4.20 | 4.27 | 4.27 | 4.33 | 4.47 | 4.60 | 4.60 | 4.60 | 4.80 | 0.27 | 7.5  |
| NaBr | 3.47 | 3.87 | 4.00 | 4.27 | 3.60 | 3.87 | 4.00 | 4.27 | 3.73 | 4.00 | 4.27 | 4.40 | 4.13 | 4.27 | 4.40 | 4.40 | 0.20 | 5.5  |
| KBr  | 3.60 | 3.73 | 4.07 | 4.47 | 3.73 | 3.87 | 4.13 | 4.47 | 4.00 | 4.07 | 4.27 | 4.53 | 4.33 | 4.40 | 4.47 | 4.67 | 0.25 | 7.0  |
| RbBr | 3.60 | 3.73 | 3.93 | 4.27 | 3.80 | 3.87 | 4.07 | 4.33 | 4.00 | 4.07 | 4.20 | 4.47 | 4.47 | 4.47 | 4.53 | 4.67 | 0.22 | 6.2  |
| CsBr | 3.80 | 3.80 | 3.93 | 4.13 | 4.00 | 4.00 | 4.07 | 4.27 | 4.20 | 4.20 | 4.27 | 4.40 | 4.60 | 4.60 | 4.60 | 4.67 | 0.29 | 8.2  |
| NaI  | 3.60 | 4.00 | 4.27 | 4.40 | 3.73 | 4.13 | 4.27 | 4.40 | 3.87 | 4.27 | 4.40 | 4.40 | 4.13 | 4.40 | 4.40 | 4.67 | 0.29 | 8.3  |
| KI   | 3.87 | 4.00 | 4.27 | 4.67 | 3.93 | 4.00 | 4.27 | 4.67 | 4.00 | 4.13 | 4.33 | 4.73 | 4.33 | 4.40 | 4.53 | 4.80 | 0.38 | 10.7 |
| RbI  | 3.87 | 3.93 | 4.13 | 4.47 | 3.93 | 4.00 | 4.20 | 4.53 | 4.00 | 4.07 | 4.27 | 4.53 | 4.33 | 4.40 | 4.53 | 4.73 | 0.32 | 9.0  |
| CsI  | 4.07 | 4.07 | 4.13 | 4.33 | 4.07 | 4.13 | 4.20 | 4.33 | 4.20 | 4.20 | 4.27 | 4.47 | 4.53 | 4.53 | 4.60 | 4.67 | 0.37 | 10.6 |
| MAE  | 0.61 | 0.45 | 0.44 | 0.46 | 0.30 | 0.14 | 0.16 | 0.17 | 0.38 | 0.16 | 0.14 | 0.12 | 0.32 | 0.14 | 0.12 | 0.13 |      |      |
| MAPE | 20.9 | 13.7 | 12.6 | 12.4 | 8.4  | 3.7  | 3.9  | 4.1  | 10.3 | 3.8  | 3.3  | 2.8  | 7.7  | 3.2  | 2.7  | 2.8  |      |      |

TABLE X. Alchemical predictions of alkali halides' lattice constants [Å] in zb symmetry. Columns indicate target $AX$ and rows correspond to the reference $AX$. The values on the diagonal are pure DFT/PBE calculations.

|      | NaF  | KF   | RbF  | CsF  | NaCl | KCl  | RbCl | CsCl | NaBr | KBr  | RbBr | CsBr | NaI  | KI   | RbI  | CsI  | MAE  | MAPE |
|------|------|------|------|------|------|------|------|------|------|------|------|------|------|------|------|------|------|------|
| NaF  | 5.20 | 6.00 | 6.20 | 6.60 | 6.60 | 7.00 | 7.00 | 7.00 | 7.20 | 7.60 | 7.60 | 7.60 | 7.80 | 7.80 | 7.80 | 7.80 | 0.37 | 4.9  |
| KF   | 5.80 | 6.00 | 6.40 | 7.00 | 7.40 | 7.40 | 7.60 | 7.80 | 7.80 | 7.80 | 8.00 | 8.20 | 8.60 | 8.60 | 8.80 | 8.80 | 0.48 | 7.1  |
| RbF  | 6.00 | 6.00 | 6.20 | 6.80 | 7.60 | 7.60 | 7.80 | 7.80 | 8.20 | 8.20 | 8.40 | 8.40 | 9.00 | 9.00 | 9.00 | 9.00 | 0.68 | 9.8  |
| CsF  | 6.40 | 6.40 | 6.40 | 6.60 | 8.00 | 8.00 | 8.00 | 8.00 | 8.60 | 8.60 | 8.60 | 8.60 | 9.40 | 9.40 | 9.40 | 9.40 | 1.00 | 14.4 |
| NaCl | 5.80 | 7.00 | 7.20 | 7.60 | 6.20 | 7.00 | 7.40 | 7.60 | 6.60 | 7.20 | 7.60 | 7.60 | 7.40 | 7.60 | 7.80 | 8.20 | 0.44 | 6.4  |
| KCl  | 6.20 | 6.60 | 7.20 | 8.00 | 6.80 | 7.00 | 7.40 | 8.00 | 7.40 | 7.40 | 7.80 | 8.20 | 8.00 | 8.20 | 8.40 | 8.60 | 0.45 | 7.1  |
| RbCl | 6.20 | 6.40 | 7.00 | 7.60 | 7.00 | 7.20 | 7.40 | 8.00 | 7.60 | 7.60 | 7.80 | 8.20 | 8.40 | 8.40 | 8.40 | 8.80 | 0.48 | 7.5  |
| CsCl | 6.80 | 6.80 | 7.00 | 7.20 | 7.40 | 7.40 | 7.60 | 7.80 | 8.00 | 8.00 | 8.00 | 8.20 | 8.80 | 8.80 | 8.80 | 8.80 | 0.71 | 10.8 |
| NaBr | 6.20 | 7.00 | 7.60 | 8.20 | 6.40 | 7.20 | 7.60 | 8.20 | 6.60 | 7.60 | 7.60 | 8.20 | 7.20 | 7.60 | 8.20 | 8.20 | 0.51 | 7.7  |
| KBr  | 6.60 | 7.00 | 7.60 | 8.40 | 6.80 | 7.20 | 7.60 | 8.40 | 7.20 | 7.40 | 7.80 | 8.40 | 8.00 | 8.00 | 8.40 | 8.80 | 0.59 | 9.3  |
| RbBr | 6.60 | 6.80 | 7.20 | 8.00 | 7.00 | 7.20 | 7.40 | 8.00 | 7.40 | 7.60 | 7.80 | 8.20 | 8.20 | 8.20 | 8.40 | 8.80 | 0.53 | 8.5  |
| CsBr | 7.00 | 7.00 | 7.20 | 7.60 | 7.40 | 7.40 | 7.60 | 7.80 | 7.80 | 7.80 | 8.00 | 8.20 | 8.60 | 8.60 | 8.60 | 8.80 | 0.71 | 11.1 |
| NaI  | 6.60 | 7.60 | 8.20 | 8.40 | 6.60 | 7.60 | 8.20 | 8.40 | 6.80 | 7.60 | 8.20 | 8.40 | 7.20 | 7.80 | 8.20 | 8.80 | 0.71 | 11.0 |
| KI   | 7.00 | 7.40 | 8.00 | 8.80 | 7.00 | 7.40 | 8.00 | 8.80 | 7.20 | 7.60 | 8.20 | 9.00 | 7.80 | 8.00 | 8.40 | 9.00 | 0.85 | 13.1 |
| RbI  | 7.00 | 7.20 | 7.60 | 8.40 | 7.20 | 7.40 | 7.80 | 8.40 | 7.40 | 7.40 | 8.00 | 8.60 | 8.00 | 8.00 | 8.40 | 8.80 | 0.72 | 11.3 |
| CsI  | 7.40 | 7.40 | 7.60 | 8.00 | 7.60 | 7.60 | 7.80 | 8.20 | 7.80 | 7.80 | 8.00 | 8.20 | 8.40 | 8.40 | 8.40 | 8.80 | 0.84 | 13.3 |
| MAE  | 1.31 | 0.84 | 1.03 | 1.17 | 0.92 | 0.37 | 0.31 | 0.36 | 0.93 | 0.35 | 0.25 | 0.23 | 1.04 | 0.45 | 0.28 | 0.23 |      |      |
| MAPE | 25.1 | 14.0 | 16.6 | 17.8 | 14.8 | 5.3  | 4.1  | 4.6  | 14.1 | 4.7  | 3.2  | 2.8  | 14.4 | 5.7  | 3.3  | 2.6  |      |      |

TABLE XI. Alchemical predictions of alkali halides' bulk modulus [$10^{11} \frac{\text{dynes}}{\text{cm}^2}$] in rs symmetry. Columns indicate target $AX$ and rows correspond to the reference $AX$. The values on the diagonal are pure DFT/PBE calculations.

|      | NaF   | KF    | RbF   | CsF   | NaCl  | KCl   | RbCl  | CsCl  | NaBr  | KBr   | RbBr  | CsBr  | NaI   | KI    | RbI   | CsI   | MAE   | MAPE |
|------|-------|-------|-------|-------|-------|-------|-------|-------|-------|-------|-------|-------|-------|-------|-------|-------|-------|------|
| NaF  | 4.075 | 3.014 | 2.727 | 2.947 | 1.642 | 1.561 | 1.414 | 1.324 | 1.698 | 1.909 | 1.954 | 2.339 | 3.128 | 3.338 | 3.384 | 3.769 | 0.98  | 83.4 |
| KF   | 3.423 | 2.978 | 2.529 | 2.238 | 1.194 | 0.886 | 1.133 | 1.031 | 1.789 | 1.819 | 2.009 | 1.024 | 1.892 | 1.902 | 1.993 | 0.872 | 0.52  | 36.9 |
| RbF  | 3.235 | 2.527 | 2.271 | 1.963 | 0.816 | 0.879 | 1.032 | 1.412 | 0.861 | 0.540 | 0.584 | 0.732 | 0.000 | 0.643 | 0.000 | 0.435 | 0.67  | 44.4 |
| CsF  | 3.006 | 2.304 | 2.729 | 1.873 | 1.077 | 1.095 | 1.159 | 0.801 | 0.633 | 0.638 | 0.657 | 0.711 | 0.000 | 0.000 | 0.455 | 0.000 | 0.76  | 49.1 |
| NaCl | 2.827 | 1.980 | 1.392 | 0.551 | 2.290 | 1.636 | 1.409 | 0.880 | 1.744 | 1.561 | 1.671 | 1.657 | 1.480 | 0.865 | 1.882 | 2.443 | 0.64  | 42.5 |
| KCl  | 2.337 | 1.923 | 1.400 | 0.953 | 1.729 | 1.608 | 1.276 | 1.167 | 1.385 | 1.210 | 1.330 | 1.102 | 1.146 | 0.856 | 1.189 | 0.807 | 0.50  | 24.5 |
| RbCl | 3.133 | 2.374 | 1.564 | 1.061 | 1.834 | 1.651 | 1.167 | 1.076 | 1.302 | 0.941 | 1.496 | 1.084 | 0.731 | 0.816 | 1.003 | 0.835 | 0.44  | 24.1 |
| CsCl | 4.609 | 3.004 | 2.340 | 1.594 | 1.601 | 1.630 | 1.383 | 1.331 | 0.997 | 1.053 | 1.144 | 1.038 | 0.953 | 0.971 | 0.465 | 0.728 | 0.30  | 18.6 |
| NaBr | 2.266 | 1.579 | 0.837 | 1.633 | 2.115 | 1.416 | 0.901 | 1.712 | 2.068 | 1.515 | 1.107 | 1.903 | 1.445 | 1.215 | 1.710 | 0.612 | 0.58  | 34.2 |
| KBr  | 2.379 | 1.705 | 1.273 | 1.250 | 1.778 | 1.620 | 1.036 | 0.521 | 1.649 | 1.208 | 1.276 | 0.912 | 0.920 | 1.077 | 1.253 | 0.839 | 0.52  | 26.8 |
| RbBr | 2.199 | 2.029 | 1.402 | 1.520 | 1.743 | 1.628 | 1.280 | 0.718 | 1.427 | 1.108 | 1.233 | 1.310 | 0.908 | 1.087 | 0.759 | 0.864 | 0.48  | 23.4 |
| CsBr | 4.407 | 2.787 | 1.887 | 1.182 | 1.479 | 1.525 | 1.403 | 1.140 | 1.201 | 1.221 | 1.099 | 1.041 | 0.556 | 0.616 | 0.768 | 0.523 | 0.40  | 24.3 |
| NaI  | 1.780 | 0.942 | 1.137 | 0.932 | 1.927 | 0.972 | 1.157 | 0.984 | 1.708 | 1.096 | 1.264 | 1.000 | 1.434 | 1.587 | 0.677 | 1.274 | 0.61  | 28.0 |
| KI   | 2.095 | 1.399 | 1.130 | 0.679 | 1.730 | 1.467 | 1.200 | 0.691 | 1.470 | 1.297 | 0.864 | 0.751 | 1.149 | 1.104 | 0.922 | 0.537 | 0.64  | 31.5 |
| RbI  | 2.257 | 1.799 | 1.088 | 1.104 | 2.415 | 1.513 | 1.250 | 1.170 | 1.892 | 1.232 | 0.984 | 0.846 | 0.846 | 1.234 | 0.851 | 0.865 | 0.47  | 22.1 |
| CsI  | 2.120 | 1.901 | 1.461 | 0.877 | 1.762 | 1.898 | 1.651 | 0.993 | 1.504 | 1.580 | 1.338 | 1.206 | 1.035 | 1.095 | 0.907 | 1.100 | 0.54  | 26.6 |
| MAE  | 1.386 | 0.902 | 0.777 | 0.711 | 0.651 | 0.238 | 0.156 | 0.351 | 0.651 | 0.290 | 0.311 | 0.309 | 0.649 | 0.451 | 0.571 | 0.631 |       |      |
| MAPE | 34.0  | 30.3  | 34.2  | 38.0  | 28.4  | 14.8  | 13.4  | 26.4  | 31.5  | 24.0  | 25.2  | 29.6  | 45.3  | 40.9  | 67.1  | 57.4  |       |      |

TABLE XII. Alchemical predictions of alkali halides' bulk modulus [$10^{11} \frac{\text{dynes}}{\text{cm}^2}$] in cscl symmetry. Columns indicate target $AX$ and rows correspond to the reference $AX$. The values on the diagonal are pure DFT/PBE calculations.

|      | NaF   | KF    | RbF   | CsF   | NaCl  | KCl   | RbCl  | CsCl  | NaBr  | KBr   | RbBr  | CsBr  | NaI   | KI    | RbI   | CsI   | MAE   | MAPE |
|------|-------|-------|-------|-------|-------|-------|-------|-------|-------|-------|-------|-------|-------|-------|-------|-------|-------|------|
| NaF  | 4.278 | 2.952 | 2.707 | 3.349 | 2.186 | 2.257 | 1.870 | 1.914 | 1.666 | 1.968 | 2.155 | 1.030 | 1.492 | 0.551 | 0.643 | 0.740 | 0.41  | 26.0 |
| KF   | 3.707 | 3.366 | 3.282 | 2.498 | 1.028 | 1.340 | 1.216 | 0.931 | 2.089 | 2.026 | 2.122 | 1.700 | 2.759 | 2.727 | 2.816 | 3.053 | 0.83  | 61.2 |
| RbF  | 3.079 | 2.835 | 2.772 | 2.315 | 1.181 | 1.210 | 1.170 | 1.196 | 0.892 | 0.926 | 0.950 | 0.859 | 0.646 | 0.667 | 0.702 | 0.426 | 0.61  | 34.5 |
| CsF  | 2.941 | 3.123 | 2.992 | 2.279 | 1.094 | 1.104 | 0.996 | 1.139 | 0.570 | 0.803 | 0.546 | 0.666 | 0.384 | 0.373 | 0.457 | 0.395 | 0.76  | 45.3 |
| NaCl | 2.943 | 2.010 | 2.263 | 1.904 | 2.392 | 1.008 | 2.359 | 3.686 | 2.479 | 2.321 | 2.758 | 6.584 | 1.550 | 1.461 | 1.559 | 1.353 | 1.09  | 72.2 |
| KCl  | 2.566 | 2.425 | 1.779 | 1.251 | 2.019 | 1.709 | 1.501 | 1.233 | 1.605 | 1.517 | 1.387 | 1.134 | 1.167 | 1.207 | 1.140 | 0.904 | 0.45  | 17.9 |
| RbCl | 2.797 | 2.535 | 2.176 | 1.404 | 1.943 | 1.825 | 1.650 | 1.329 | 1.384 | 1.424 | 1.438 | 1.208 | 1.134 | 0.992 | 1.076 | 0.943 | 0.41  | 16.8 |
| CsCl | 3.600 | 2.421 | 2.651 | 1.700 | 1.690 | 1.676 | 1.579 | 1.444 | 1.288 | 1.296 | 1.212 | 1.079 | 1.025 | 1.027 | 1.059 | 0.842 | 0.36  | 16.7 |
| NaBr | 2.607 | 1.791 | 1.588 | 1.301 | 2.195 | 1.941 | 1.669 | 1.436 | 2.204 | 1.716 | 0.990 | 0.958 | 1.541 | 1.520 | 1.193 | 1.838 | 0.54  | 26.1 |
| KBr  | 1.946 | 1.914 | 1.441 | 1.059 | 2.177 | 1.824 | 1.410 | 1.091 | 1.587 | 1.497 | 1.262 | 1.041 | 1.262 | 1.186 | 1.165 | 1.169 | 0.58  | 23.7 |
| RbBr | 2.341 | 1.900 | 1.645 | 1.278 | 1.975 | 1.846 | 1.525 | 1.256 | 1.638 | 1.526 | 1.387 | 1.085 | 1.006 | 1.114 | 1.071 | 1.001 | 0.52  | 20.1 |
| CsBr | 2.273 | 2.277 | 1.826 | 1.444 | 1.731 | 1.761 | 1.650 | 1.323 | 1.408 | 1.408 | 1.328 | 1.187 | 0.969 | 1.008 | 1.180 | 1.157 | 0.51  | 21.3 |
| NaI  | 2.617 | 1.684 | 1.421 | 1.571 | 2.010 | 1.386 | 1.405 | 1.437 | 1.808 | 1.250 | 1.170 | 1.448 | 1.531 | 1.136 | 1.477 | 0.419 | 0.57  | 26.5 |
| KI   | 2.112 | 1.539 | 1.268 | 0.977 | 1.903 | 1.638 | 1.316 | 1.035 | 1.820 | 1.429 | 1.231 | 0.808 | 1.276 | 1.175 | 1.104 | 1.062 | 0.63  | 25.8 |
| RbI  | 1.929 | 1.733 | 1.436 | 1.242 | 1.830 | 1.658 | 1.350 | 0.966 | 1.823 | 1.579 | 1.311 | 1.208 | 1.265 | 1.187 | 1.015 | 0.884 | 0.58  | 22.2 |
| CsI  | 1.722 | 1.737 | 1.516 | 1.235 | 1.895 | 1.931 | 1.643 | 1.326 | 1.566 | 1.394 | 1.320 | 1.093 | 1.075 | 1.076 | 1.031 | 0.986 | 0.59  | 22.0 |
| MAE  | 1.666 | 1.174 | 0.870 | 0.820 | 0.602 | 0.272 | 0.266 | 0.386 | 0.666 | 0.281 | 0.365 | 0.555 | 0.462 | 0.325 | 0.329 | 0.404 |       |      |
| MAPE | 38.9  | 34.9  | 31.4  | 36.0  | 25.1  | 15.9  | 16.1  | 26.7  | 30.2  | 18.8  | 26.3  | 46.8  | 30.2  | 27.6  | 32.4  | 41.0  |       |      |

TABLE XIII. Alchemical predictions of alkali halides' bulk modulus [$10^{11} \frac{\text{dynes}}{\text{cm}^2}$] in zb symmetry. Columns indicate target $AX$ and rows correspond to the reference $AX$. The values on the diagonal are pure DFT/PBE calculations.

|      | NaF   | KF    | RbF   | CsF   | NaCl  | KCl   | RbCl  | CsCl  | NaBr  | KBr   | RbBr  | CsBr  | NaI   | KI    | RbI   | CsI   | MAE  | MAPE |
|------|-------|-------|-------|-------|-------|-------|-------|-------|-------|-------|-------|-------|-------|-------|-------|-------|------|------|
| NaF  | 3.135 | 2.013 | 1.763 | 1.480 | 1.171 | 0.969 | 1.074 | 1.301 | 0.800 | 0.412 | 0.542 | 0.699 | 0.653 | 0.764 | 0.863 | 0.965 | 0.26 | 28.9 |
| KF   | 2.448 | 1.957 | 1.552 | 1.166 | 0.742 | 0.798 | 0.669 | 0.672 | 0.824 | 0.869 | 0.660 | 0.589 | 0.452 | 0.462 | 0.290 | 0.402 | 0.35 | 30.3 |
| RbF  | 1.979 | 1.973 | 1.987 | 1.320 | 0.842 | 0.880 | 0.642 | 0.857 | 0.612 | 0.627 | 0.447 | 0.526 | 0.321 | 0.352 | 0.343 | 0.398 | 0.39 | 33.7 |
| CsF  | 1.507 | 1.539 | 1.699 | 1.513 | 0.722 | 0.729 | 0.759 | 0.834 | 0.501 | 0.503 | 0.513 | 0.539 | 0.248 | 0.249 | 0.246 | 0.000 | 0.52 | 45.2 |
| NaCl | 1.984 | 1.050 | 1.012 | 0.978 | 1.357 | 1.247 | 0.920 | 1.058 | 1.417 | 1.089 | 0.884 | 1.143 | 0.927 | 0.954 | 0.000 | 0.491 | 0.36 | 28.1 |
| KCl  | 1.842 | 1.325 | 1.006 | 0.648 | 1.315 | 1.191 | 0.981 | 0.748 | 0.901 | 0.926 | 0.000 | 0.846 | 0.770 | 0.575 | 0.533 | 0.668 | 0.42 | 31.4 |
| RbCl | 1.599 | 1.472 | 1.087 | 0.888 | 1.186 | 1.029 | 0.964 | 0.643 | 0.822 | 0.871 | 0.870 | 0.741 | 0.546 | 0.611 | 0.767 | 0.443 | 0.38 | 24.4 |
| CsCl | 1.707 | 1.728 | 1.343 | 1.137 | 1.015 | 1.012 | 0.886 | 0.822 | 0.685 | 0.721 | 0.862 | 0.795 | 0.400 | 0.440 | 0.487 | 0.658 | 0.38 | 28.5 |
| NaBr | 1.474 | 1.173 | 0.906 | 0.466 | 1.452 | 1.073 | 0.958 | 0.473 | 1.411 | 0.614 | 1.143 | 0.560 | 1.059 | 1.159 | 0.489 | 0.831 | 0.46 | 34.3 |
| KBr  | 1.455 | 1.026 | 0.808 | 0.597 | 1.370 | 0.999 | 1.030 | 0.658 | 1.044 | 0.998 | 0.981 | 0.803 | 0.622 | 0.852 | 0.618 | 0.505 | 0.42 | 25.5 |
| RbBr | 1.648 | 1.290 | 1.022 | 0.712 | 1.232 | 1.038 | 1.108 | 1.022 | 0.953 | 0.784 | 0.874 | 0.715 | 0.649 | 0.638 | 0.612 | 0.463 | 0.39 | 24.5 |
| CsBr | 1.469 | 1.442 | 1.208 | 0.969 | 1.130 | 1.159 | 0.958 | 0.853 | 0.899 | 0.946 | 0.749 | 0.680 | 0.538 | 0.554 | 0.574 | 0.531 | 0.36 | 22.2 |
| NaI  | 1.251 | 1.009 | 0.626 | 0.911 | 1.348 | 1.035 | 0.627 | 0.973 | 1.215 | 1.140 | 0.675 | 0.868 | 1.043 | 1.102 | 0.879 | 0.453 | 0.45 | 29.8 |
| KI   | 1.287 | 0.844 | 0.769 | 0.559 | 1.343 | 0.952 | 0.832 | 0.578 | 1.170 | 0.814 | 0.641 | 0.000 | 0.761 | 0.769 | 0.722 | 0.563 | 0.50 | 34.9 |
| RbI  | 1.380 | 1.103 | 1.044 | 0.779 | 1.179 | 0.940 | 0.815 | 0.836 | 0.980 | 1.147 | 0.681 | 0.660 | 0.684 | 0.886 | 0.648 | 0.582 | 0.42 | 25.9 |
| CsI  | 1.441 | 1.450 | 1.257 | 0.797 | 1.218 | 1.254 | 1.010 | 0.628 | 1.035 | 1.077 | 0.855 | 0.856 | 0.683 | 0.711 | 0.802 | 0.479 | 0.35 | 22.4 |
| MAE  | 1.504 | 0.604 | 0.847 | 0.619 | 0.220 | 0.199 | 0.130 | 0.167 | 0.488 | 0.223 | 0.225 | 0.170 | 0.424 | 0.229 | 0.205 | 0.147 |      |      |
| MAPE | 48.0  | 30.9  | 42.6  | 40.9  | 16.2  | 16.7  | 13.5  | 20.4  | 34.6  | 22.4  | 25.8  | 25.0  | 40.7  | 29.8  | 31.7  | 30.6  |      |      |